\documentclass[11pt]{article}
\usepackage{dcolumn}
\usepackage{bm}
\usepackage{booktabs}
\usepackage{graphics,subfigure}
\usepackage{epsfig}
\usepackage{color}
\usepackage{amsmath,amssymb}
\usepackage{cite}

\allowdisplaybreaks[3]

\newcommand{\Backgroundnu}[1][{}]{\ensuremath{e^+e^-  \to \nu \bar\nu \gamma}#1}

\newcommand{\signal}[1][{}]{\ensuremath{e^+e^-\rightarrow \tilde\chi^0_1 \tilde\chi^0_1\gamma}#1}

\newcommand{\GeV}{\ensuremath{~\mathrm{GeV}}}
\newcommand{\fb}{\ensuremath{~\mathrm{fb}}}

\def\lsim{\raise0.3ex\hbox{$\;<$\kern-0.75em\raise-1.1ex\hbox{$\sim\;$}}}
\def\gsim{\raise0.3ex\hbox{$\;>$\kern-0.75em\raise-1.1ex\hbox{$\sim\;$}}}

\begin{document}
\title{\hfill \phantom{Hallo}\\[-30mm]
  \hfill\mbox{\small BONN-TH-2007-01}\\[-3mm]
  \hfill\mbox{\small hep-ph/0703009}\\[5mm]
The Role of Beam polarization for Radiative Neutralino Production at the ILC}
\date{\small Physikalisches Institut der Universit\"at Bonn\\Nu\ss allee 12, 53115 Bonn}
\author{Herbert K. Dreiner, Olaf Kittel, Ulrich Langenfeld}
\maketitle
\begin{abstract}

We analyze the impact of electron and positron beam polarization on
radiative neutralino production at the International Linear Collider
(ILC).  We focus on three different mSUGRA scenarios in turn at the
Higgs strahlung threshold, the top pair production threshold, and at
$\sqrt{s} =500\GeV$. In these scenarios at the corresponding
$\sqrt{s}$, radiative neutralino production is the only supersymmetric
production mechanism which is kinematically allowed. The heavier
neutralinos, and charginos as well as the sleptons, squarks and
gluinos are too heavy to be pair produced. We calculate the signal
cross section and also the Standard Model background from radiative
neutrino production. For our scenarios, we obtain significances larger
than $10$ and signal to background ratios between $2\%$ and $5\%$, if
we have electron beam polarization $P_{e^-} = 0.0- 0.8$ and positron
beam polarization $P_{e^+} = 0.0 - 0.3$. If we have electron beam
polarization of $P_{e^-} = 0.9$, then the signal is observable with
$P_{e^+} = 0.0$ but both the significance and the signal to background
ratio are significantly improved for $P_{e^+} = 0.3$.
\end{abstract}

\section{Introduction}

The minimal supersymmetric Standard Model (MSSM) is a promising
extension of the Standard Model of particle physics (SM)~\cite{
nilles,Haber:1984rc,drees,aitchison}. If weak scale supersymmetry
(SUSY) exists, it should be discovered at the LHC~\cite{
Weiglein:2004hn}, which is scheduled to start in 2008.
However, detailed measurements of the masses, decay widths, couplings,
and spins of the discovered particles are only possible at the
international linear collider (ILC)~\cite{Weiglein:2004hn,Djouadi:2007ik,
Aguilar-Saavedra:2001rg, Abe:2001nn,Abe:2001gc}. In the first stage of
the ILC, the center-of-mass energy will be $\sqrt{s} = 500\GeV$ and
the luminosity, $\mathcal{L}$, will be $500\fb^{-1}$
within the first four years.

\medskip

In preparing for the ILC, there is an on-going debate over the extent
of beam polarization to be included in the initial 
design~\cite{Djouadi:2007ik,talk1,
Moortgat-Pick:2005cw,talk2,talk3}. It is clear that there will be at
least 80\% polarization of the electron beam, possibly even 
90\%~\cite{TDR}. A polarized positron beam is technically and financially
more involved. However, it is possible to achieve 30\% polarization
through the undulator based production of the positrons,
already from the start~\cite{talk3}. Depending on the undulator length,
even a higher degree of positron polarization
can be achieved.
In light of this discussion, it is the purpose of
this letter to reconsider the effect of various degrees of electron
and positron polarization on a particular supersymmetric production
process, namely the radiative production of the lightest neutralino
mass eigenstate $\tilde\chi^0_1$
\begin{equation}
e^+ + e^-\rightarrow \tilde\chi^0_1+\tilde\chi_1^0+\gamma.
\label{eeXXg}
\end{equation}
We shall focus on specific regions of the supersymmetric parameter
space. The signal is a single, highly energetic photon and missing
energy, carried by the neutralinos.

\medskip

The process~(\ref{eeXXg}) was previously studied within the MSSM and
with general neutralino mixing in Refs.~\cite{Datta:1994ac,
  Datta:1996ur,Ambrosanio:1995it}. The additional effect
of polarized beams was considered in Refs.~\cite{Choi:1999bs,
  Baer:2001ia,Dreiner:2006sb}. In Ref.~\cite{Dreiner:2006sb}, it was
shown that polarized beams significantly enhance the signal and
simultaneously suppress the main SM photon background from radiative
neutrino production, 
\begin{equation}
e^++e^- \to \nu +\bar\nu +\gamma\,.
\label{eenng}
\end{equation} 
Moreover, it was pointed out that for certain regions of the MSSM
parameter space, the process~(\ref{eeXXg}) is kinematically the
\textit{only} accessible SUSY production mechanism in the first stage
of the ILC at $\sqrt s=500\GeV$~\cite{Dreiner:2006sb}. Here the
heavier electroweak gauginos and the sleptons are too heavy to be pair
produced, \textit{i.e.} their masses are above 250 GeV.

\medskip

Other than the standard center-of-mass energy, $\sqrt{s}=500\GeV$, at
the ILC, also lower energies are of particular interest, namely for
Higgs and top physics. Higgs strahlung,
\begin{equation}
  e^++e^- \rightarrow Z +h,
\end{equation}
can be well studied at the threshold energy $\sqrt{s} = m_h + m_Z$,
which is $\sqrt{s}\approx220\GeV$, for a Higgs boson mass of $m_h
\approx 130\GeV$. The CP-quantum number and the spin of the Higgs
boson can be determined from an energy scan of the production cross
section near the threshold~\cite{Dova:2003py}.

\medskip

From a scan at the threshold energy of top pair production,
$\sqrt{s}=2m_t\approx350\GeV$, the top mass $m_t$ can be determined
with an error $\delta m_t<0.1\GeV$~\cite{Djouadi:2007ik}. Thus the present
error on the top mass, $\delta m_t\approx 3\GeV$~\cite{Yao:2006px},
and the foreseen error from LHC measurements, $\delta m_t\approx1\GeV
$~\cite{Borjanovic:2004ce}, can be reduced by one order of magnitude.
Also the top width, $\Gamma_t$, and the strong coupling constant,
$\alpha_s$, can be precisely determined by a multi parameter fit of
the cross section, top momentum distribution, and forward-backward
charge asymmetry near threshold~\cite{Martinez:2002st}.

\medskip

In this letter, we take these physics questions as a motivation to
study the role of polarized beams in radiative neutralino production
at the energies $\sqrt{s} = 220\GeV$, $350\GeV$, and $500\GeV$ at the
ILC. For each beam energy, we shall focus on a specific supersymmetric
parameter set within the context of minimal supergravity grand
unification (mSUGRA)~\cite{msugra}. We thus consider three mSUGRA
scenarios, which we label A, B and C, respectively, and which are
listed below in Table~\ref{tab:scenarios} together with the resulting
spectra in Table~\ref{tab:masses}. We restrict ourselves to mSUGRA
scenarios, in order to reduce the number of free parameters and since
we find it sufficient to illustrate our point. The specific scenarios
are chosen such that radiative neutralino production is the
\textit{only} supersymmetric production mechanism which is
kinematically accessible at the given center-of-mass energy.  It is
thus of particular interest to learn as much about supersymmetry as is
possible through this mechanism. As we shall see, beam polarization is
very helpful in this respect.

\medskip

In Sect.~\ref{definitions}, we define the significance, the signal to
background ratio and define a first set of experimental cuts. In 
Sect.~\ref{numerics}, we study numerically the dependence of the signal cross
sec\-tion and the SM background, the significance, and the signal to
background ratio on the beam polarization.  In particular, we compare
the results for different sets of beam polarizations, $(P_{e^+}|P_
{e^-}) = (0|0)$, $(0|0.8)$, $(-0.3|0.8)$, $(-0.6|0.8)$, $
{(0|0.9)}$ and ${(-0.3|0.9)}$.  We summarize and conclude in
Sect.~\ref{Summary}.


\section{Definitions and Cuts\label{definitions}}

We wish to study the beam polarization dependence of the cross
sections for the signal process~(\ref{eeXXg}) and the background
process~(\ref{eenng}). In the following, we define the theoretical
significance as~\cite{Dreiner:2006sb}
\begin{eqnarray}\label{significance}
S &= & \frac{\sigma_{\mathrm{Signal}}}{\sqrt{\sigma_{\mathrm{Signal}}+
\sigma_{\mathrm{Background}}}} \sqrt{\mathcal{L}},
\end{eqnarray}
and the signal to background ratio (or reliability) as~\cite{Dreiner:2006sb}
\begin{eqnarray}\label{sigbck}
r &= & \frac{\sigma_{\mathrm{Signal}}}{\sigma_{\mathrm{Background}}}\enspace .
\end{eqnarray}
Here $\sigma_{\mathrm{Signal}}$ and $\sigma_{\mathrm{Background}}$
shall refer to the signal and background cross sections also after the
cuts we impose on the final state phase space. A significance of $S=1$
implies that the signal can be measured at the statistical $68\%
$-level.  If the experimental error for the background is, for
example, $1\%$, the signal to background ratio must be greater than
$1\%$~\cite{dittmar}. For a signal to be detectable at the ILC
requires at least
\begin{equation}
S > 1 \quad \mathrm{and}\quad r>1\%\,.
\end{equation}  
We remark that these criteria can only be seen as rough estimates to
judge whether an excess of signal photons can be measured over the
background, since we do not include any detector simulation, with
effective particle reconstruction efficiencies. A detailed Monte
Carlo study is beyond the scope of the present work.

\medskip

For the tree-level calculation of the cross sections, we use the
formul{\ae} for the amplitudes squared as given in Ref.~\cite{
Dreiner:2006sb}, where details of the neutralino mixing can also be
found.  To regularize the infrared and collinear divergences of the
tree-level cross sections for signal and background, we apply cuts on
the photon scattering angle $\theta_\gamma$~\cite{Dreiner:2006sb}
\begin{equation}
 |\cos\theta_\gamma| \le 0.99\,,
\label{cuts1}
\end{equation}
and on the photon energy $E_\gamma$~\cite{Dreiner:2006sb}
\begin{equation}
0.02 \le x\le \;1-\frac{m_{\chi_1^0}^2}{E_{\rm beam}^2},
\quad \quad
x = \frac{E_\gamma}{E_{\rm beam}}, 
\label{cuts2}
\end{equation}
with the beam energy: $E_{\rm beam}=\sqrt{s}/2$.  The upper cut on the
photon energy $x^{\mathrm{max}}=1-m_{\chi_1^0}^2/E_{\rm beam}^2$ is
the kinematical limit of radiative neutralino production,
and also reduces much of the on-shell $Z$ boson contribution 
to the background
from radiative neutrino production~\cite{Dreiner:2006sb}.
We assume
that the mass of the lightest neutralino $m_{\chi_1^0}$ is known from
LHC measurements~\cite{Weiglein:2004hn}. 
Although  $m_{\chi_1^0}$ can be measured at the percent-level at the LHC,
such a precision will not be required for the upper
cut $x^{\mathrm{max}}$, since the endpoint of the photon 
energy distribution is not densely populated with signal events.
Finally, note that the ratios $S$ and $r$ only weakly  depend
on the absolute values of the cuts $|\cos\theta_\gamma| \le 0.99$, 
and $0.02 \le x$,
since signal and background have similar distributions
in energy $E_\gamma$ and angle $\theta_\gamma$, 
see Ref.~\cite{Dreiner:2006sb}.

\section{Numerical Results \label{numerics}}

We define three mSUGRA scenarios A, B, and C for the three different
energies $\sqrt{s}=220\GeV$, $350\GeV$, and $500\GeV$, respectively,
see Table~\ref{tab:scenarios}. We choose the three scenarios in such a
way, that only the lightest neutralinos can be radiatively produced
for each of the $\sqrt s$ values, respectively. The other SUSY
particles, \textit{i.e.} the heavier neutralinos and charginos, as
well as the sleptons and squarks are too heavy to be pair produced at
the ILC. It is thus of paramount interest to have an optimal
understanding of the signature~(\ref{eeXXg}), in order to learn as
much as possible about SUSY at a given ILC beam energy. Note that in
the three scenarios (A,B,C) the squark and gluino masses are below
$\{600,800,1000\}$ GeV, respectively and should be observable at the
LHC~\cite{Weiglein:2004hn}. 

\medskip

Scenario A is related to the Snowmass point SPS1a~\cite{
Aguilar-Saavedra:2005pw,Allanach:2002nj,Ghodbane:2002kg} by scaling
the common scalar mass $M_0$, the unified gaugino mass $M_ {1/2}$, and
the common trilinear coupling $A_0$ by $0.9$. Thus the slope $M_0 =
-A_0 = 0.4\, M_{1/2}$ remains unchanged. For scenarios B and C, we
also choose $M_0 = - A_0$, however we change the slopes to $M_0 =
0.42\, M_{1/2}$ in scenario B, and $M_0 = 0.48\, M_{1/2}$ in scenario
C. For all scenarios, we fix the ratio $\tan\beta = 10$ of the vacuum
expectation values of the two neutral Higgs fields. In
Table~\ref{tab:scenarios}, we explicitly give the relevant low energy
mSUGRA parameters for all scenarios. These are the $U(1)$ and $SU(2)$
gaugino mass parameters $M_1$ and $M_2$, respectively, and the
higgsino mass parameter $\mu$. The masses of the light neutralinos,
charginos, and sleptons are given in Table~\ref{tab:masses}.  All
parameters and masses are calculated at one-loop order with the
computer code SPheno~\cite{Porod:2003um}.

\medskip

\begin{center}
\begin{figure}[ht]
\setlength{\unitlength}{1cm}
\subfigure[Signal cross section $\sigma(e^+e^- \to \tilde\chi^0_1\tilde\chi^0_1\gamma)$ in fb.
\label{fig:sigma220}]%
        {\scalebox{0.53}{\includegraphics{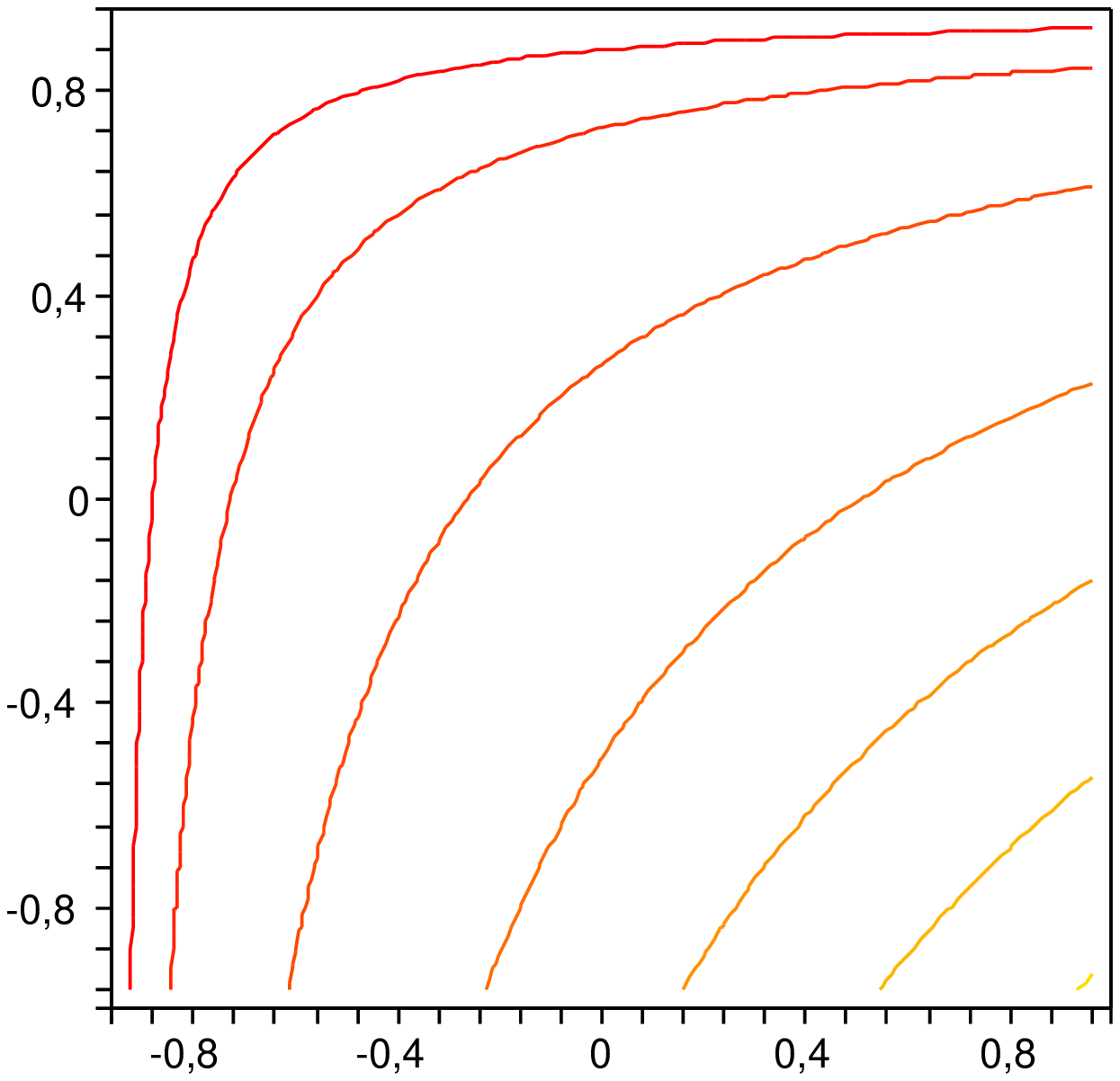}}
\put(-1.2,0.0){$P_{e^-}$}
\put(-8.0,6.5){$P_{e^+}$}
\put(-1.2,1.0){${\scriptstyle{25}}$}
\put(-1.8,1.5){${\scriptstyle{20}}$}
\put(-2.6,2.2){${\scriptstyle{15}}$}
\put(-3.4,3.1){${\scriptstyle{10}}$}
\put(-4.5,4.2){${\scriptstyle{5}}$}
\put(-5.5,5.2){${\scriptstyle{2}}$}
\put(-6.1,5.9){${\scriptstyle{1}}$}
}
\hspace{1mm}
\subfigure[Background cross section $\sigma(e^+e^- \to \nu \bar\nu \gamma)$ in fb.\label{fig:back220}]%
        {\scalebox{0.53}{\includegraphics{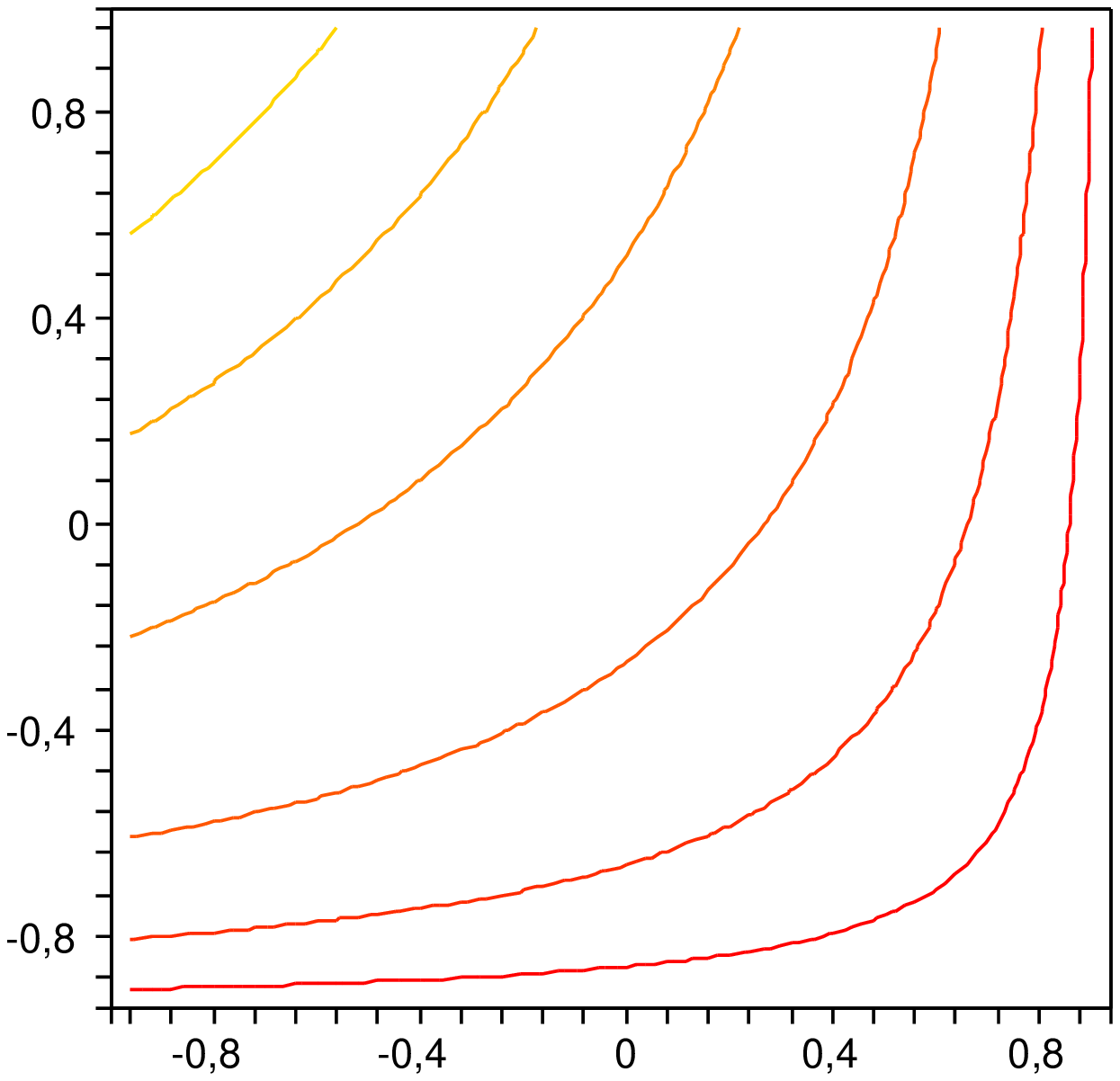}}
\put(-1.2,0){$P_{e^-}$}
\put(-8,6.5){$P_{e^+}$}
\put(-2.0,1.7){${\scriptstyle{500}}$}
\put(-2.9,2.4){${\scriptstyle{1000}}$}
\put(-3.8,3.2){${\scriptstyle{2000}}$}
\put(-4.9,4.4){${\scriptstyle{4000}}$}
\put(-5.8,5.3){${\scriptstyle{6000}}$}
\put(-6.4,6.1){${\scriptstyle{8000}}$}
}

\subfigure[Significance $S$.\label{fig:signi220}]{\scalebox{0.53}{\includegraphics{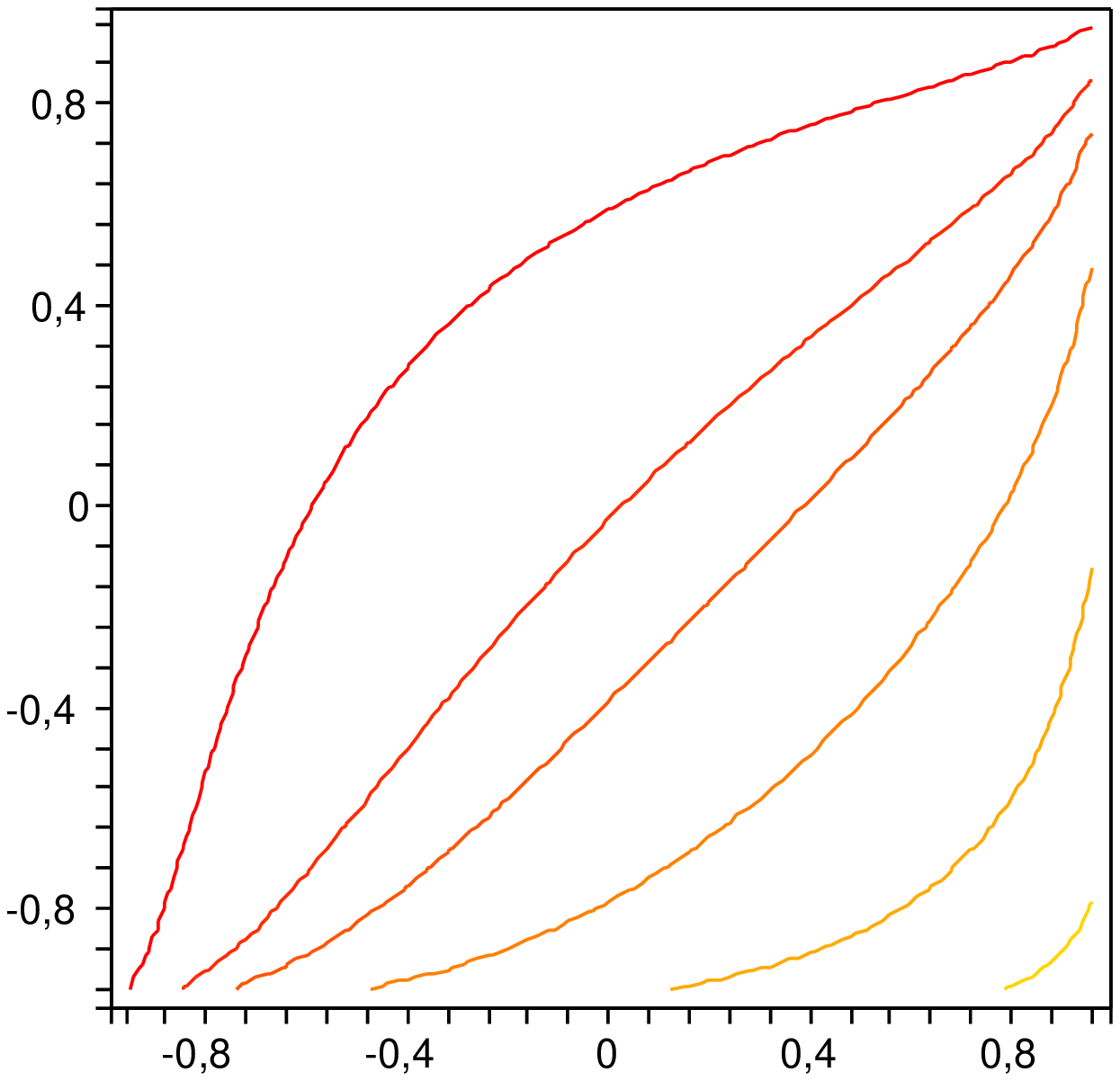}}
\put(-1.2,0){$P_{e^-}$}
\put(-8,6.5){$P_{e^+}$}
\put(-1.3,1.1){${\scriptstyle{30}}$}
\put(-2.0,1.6){${\scriptstyle{20}}$}
\put(-2.9,2.2){${\scriptstyle{10}}$}
\put(-3.7,2.8){${\scriptstyle{5}}$}
\put(-4.2,3.4){${\scriptstyle{3}}$}
\put(-5.3,4.4){${\scriptstyle{1}}$}
}
\hspace{1mm}
\subfigure[Signal to background ratio $r$ in \%. \label{fig:reli220}]{\scalebox{0.53}{\includegraphics{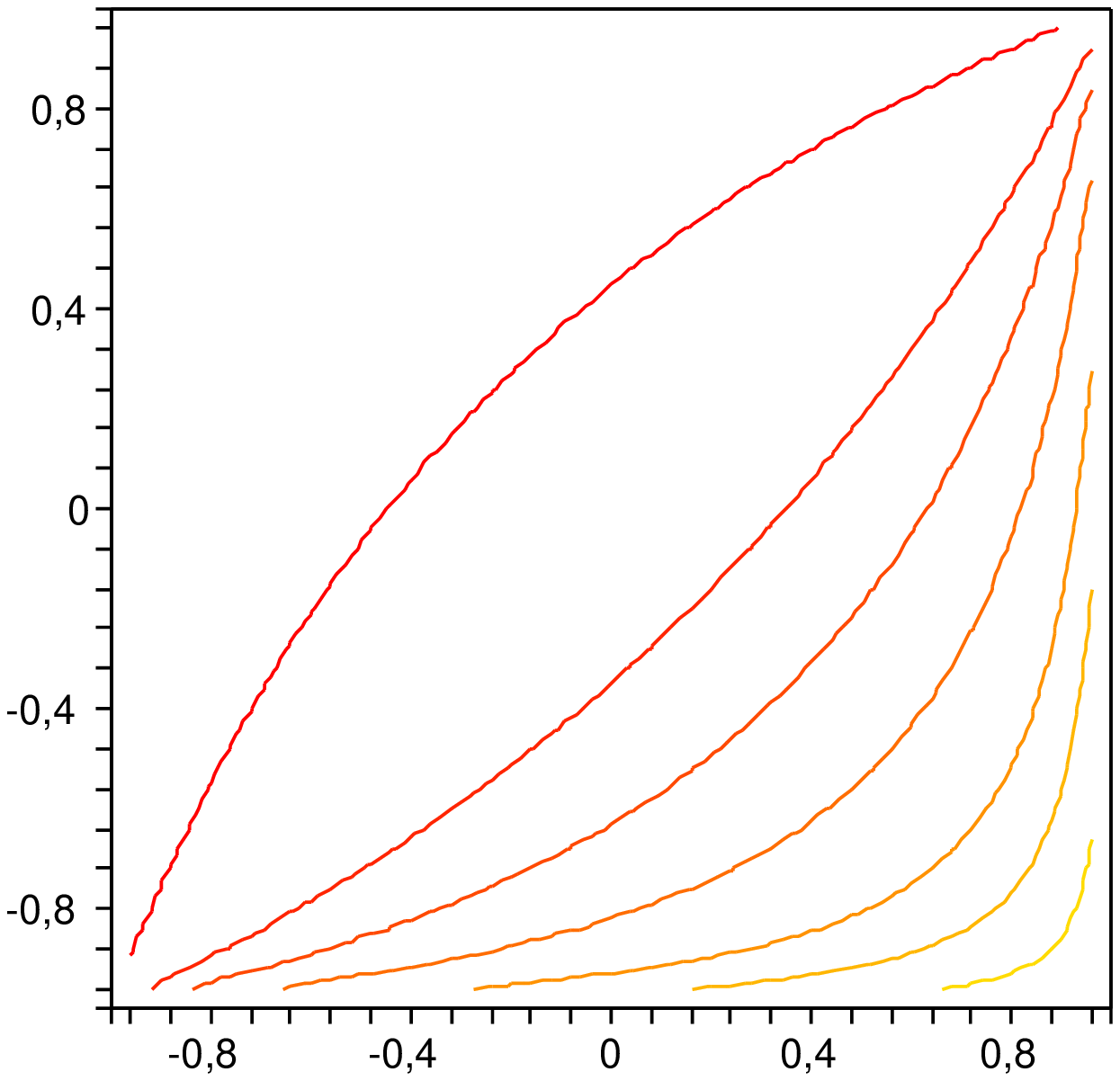}}
\put(-1.2,0){$P_{e^-}$}
\put(-8,6.5){$P_{e^+}$}
\put(-1.3,1.1){${\scriptstyle{8}}$}
\put(-1.7,1.3){${\scriptstyle{6}}$}
\put(-2.1,1.6){${\scriptstyle{4}}$}
\put(-2.6,2.0){${\scriptstyle{2}}$}
\put(-3.1,2.5){${\scriptstyle{1}}$}
\put(-3.7,3.1){${\scriptstyle{0.5}}$}
\put(-5.0,4.4){${\scriptstyle{0.1}}$}
}
\caption{Signal cross section (a), background cross section (b), 
significance (c), and signal to background ratio (d) for $\sqrt{s} =
220\GeV$, and an integrated luminosity $\mathcal{L}=500\fb^{-1}$ for
scenario A: $M_0 = 90\GeV$, $M_{1/2} = 225\GeV$, $A_0 = -90\GeV$, and
$\tan\beta = 10$, see Tables~\ref{tab:scenarios} and~\ref{tab:masses}.}
\label{fig:scenarioA}
\end{figure}
\end{center}

\begin{figure}[ht]
\setlength{\unitlength}{1cm}
\subfigure[Signal cross section $\sigma(e^+e^- \to \tilde\chi^0_1\tilde\chi^0_1\gamma)$ in fb.
\label{fig:sigma350}]{\scalebox{0.53}{\includegraphics{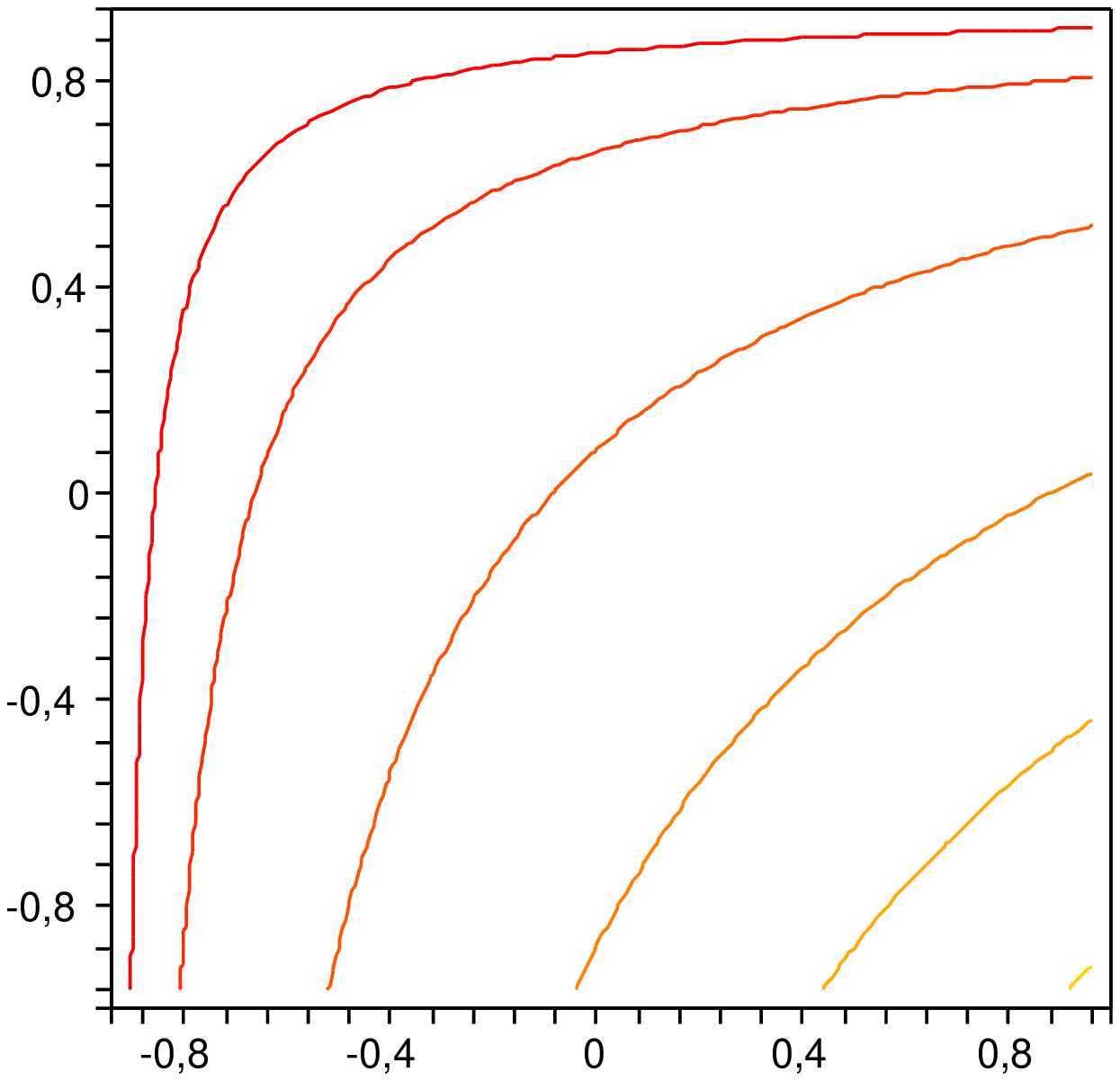}}%
\put(-1.2,0.0){$P_{e^-}$}
\put(-8.0,6.5){$P_{e^+}$}
\put(-1.2,1.0){${\scriptstyle{20}}$}
\put(-2.1,1.7){${\scriptstyle{15}}$}
\put(-3.0,2.7){${\scriptstyle{10}}$}
\put(-4.2,3.9){${\scriptstyle{5}}$}
\put(-5.4,5.1){${\scriptstyle{2}}$}
\put(-6.0,5.8){${\scriptstyle{1}}$}
}
\hspace{1mm}
\subfigure[Background cross section $\sigma(e^+e^- \to \nu \bar\nu \gamma)$ in fb.\label{fig:back350}]
        {\scalebox{0.53}{\includegraphics{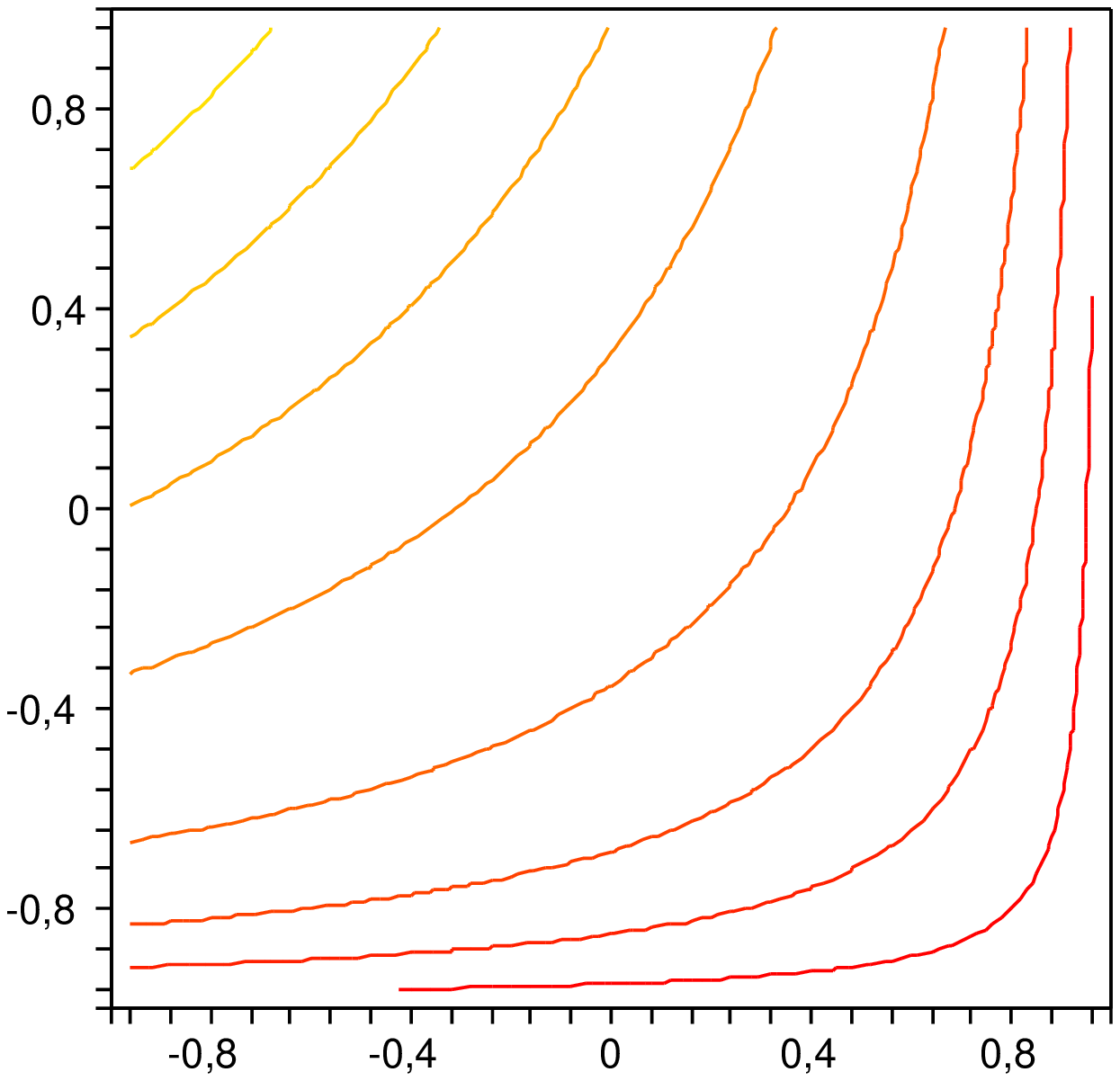}}
\put(-1.2,0){$P_{e^-}$}
\put(-8,6.5){$P_{e^+}$}
\put(-1.7,1.4){${\scriptstyle{200}}$}
\put(-2.3,2.0){${\scriptstyle{500}}$}
\put(-3.0,2.5){${\scriptstyle{1000}}$}
\put(-3.7,3.2){${\scriptstyle{2000}}$}
\put(-4.8,4.2){${\scriptstyle{4000}}$}
\put(-5.4,5.1){${\scriptstyle{6000}}$}
\put(-6.1,5.7){${\scriptstyle{8000}}$}
\put(-6.6,6.3){${\scriptstyle{10000}}$}
}

\subfigure[Significance $S$.\label{fig:signi350}]{\scalebox{0.53}{\includegraphics{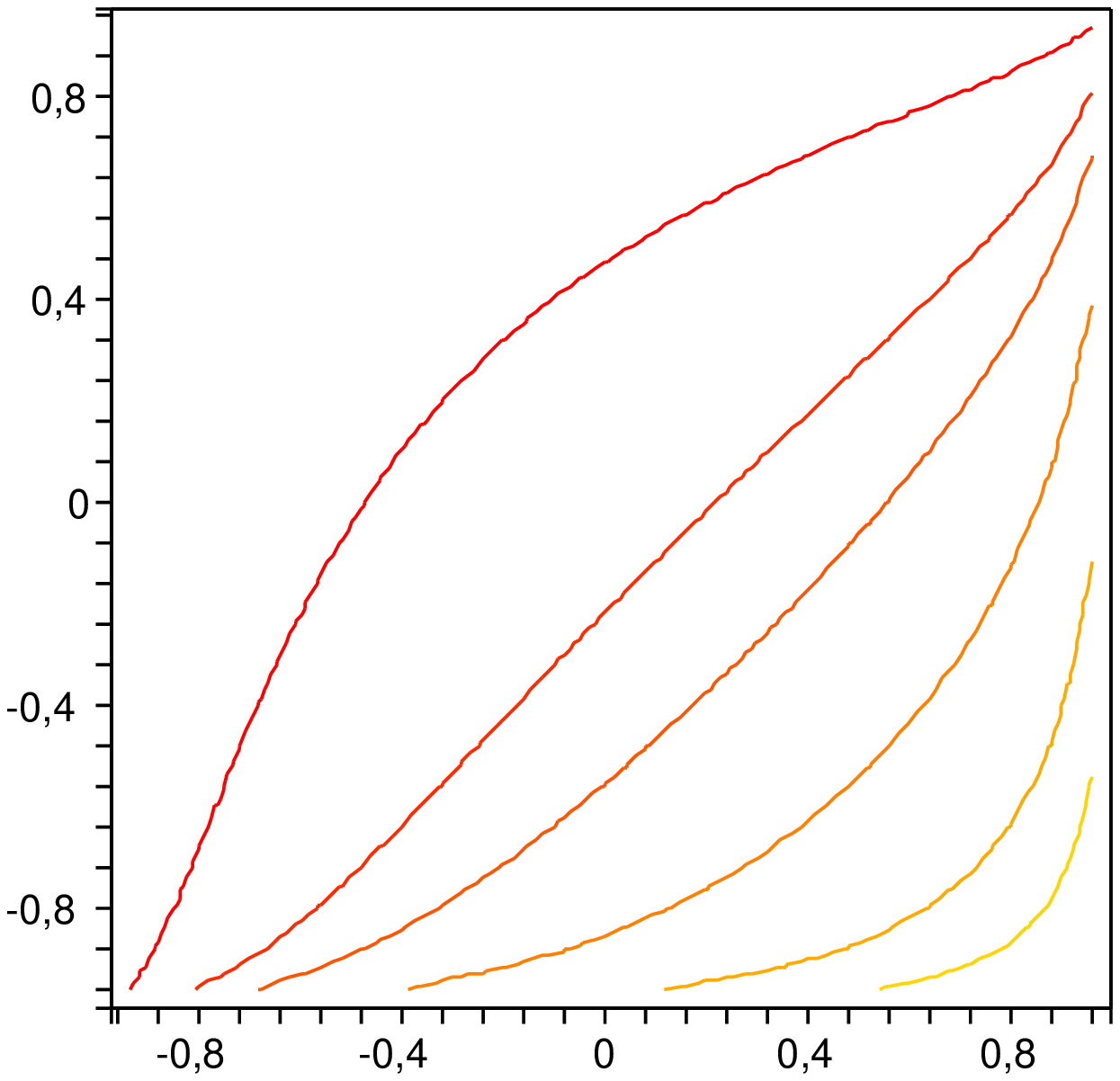}}
\put(-1.2,0){$P_{e^-}$}
\put(-8,6.5){$P_{e^+}$}
\put(-1.3,1.0){${\scriptstyle{30}}$}
\put(-1.9,1.5){${\scriptstyle{20}}$}
\put(-2.7,2.0){${\scriptstyle{10}}$}
\put(-3.4,2.6){${\scriptstyle{5}}$}
\put(-3.9,3.1){${\scriptstyle{3}}$}
\put(-5.1,4.2){${\scriptstyle{1}}$}
}
\hspace{1mm}
\subfigure[Signal to background ratio $r$ in \%.\label{fig:reli350}]{\scalebox{0.53}{\includegraphics{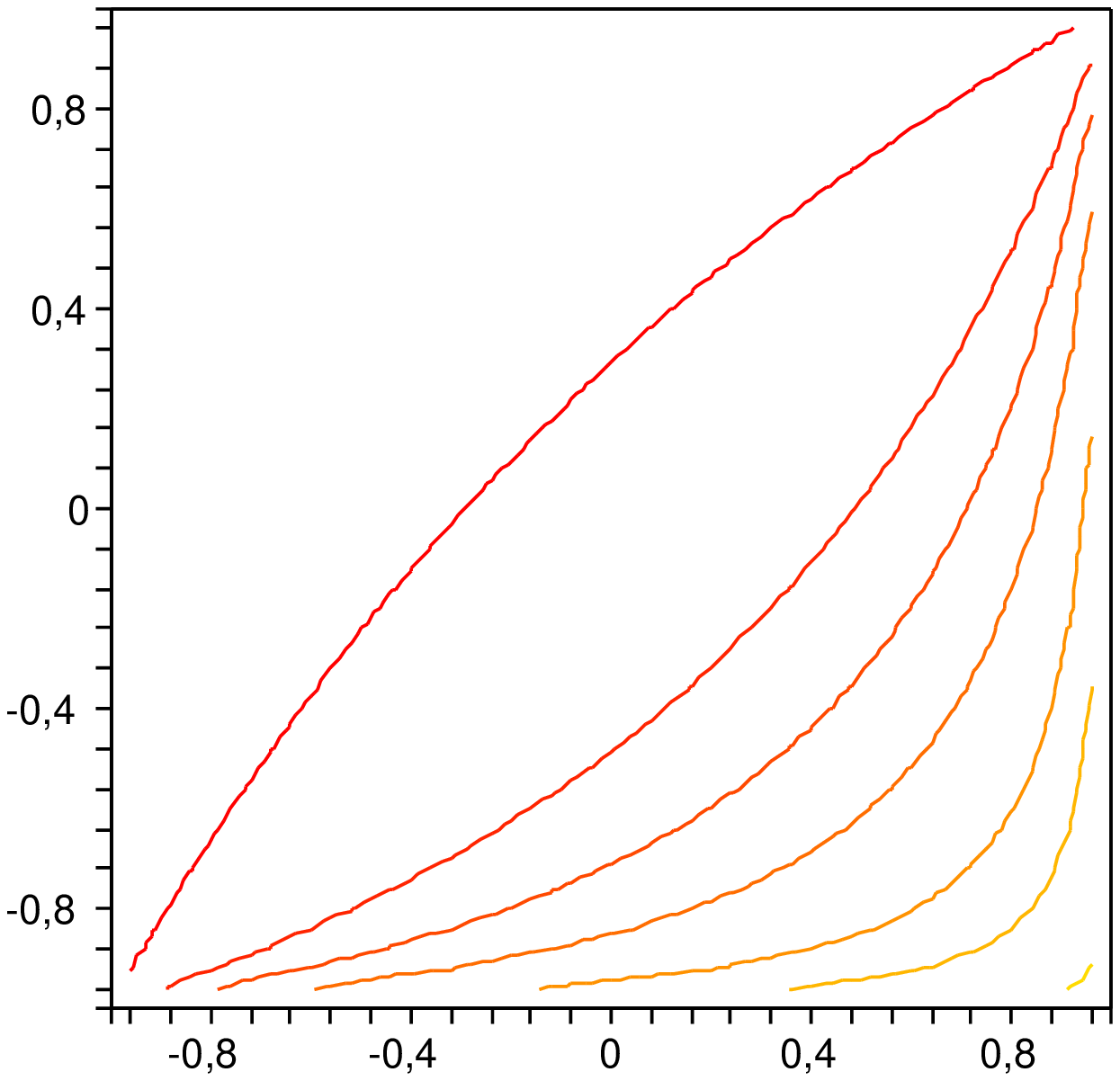}}
\put(-1.2,0){$P_{e^-}$}
\put(-8,6.5){$P_{e^+}$}
\put(-1.2,0.9){${\scriptstyle{20}}$}
\put(-1.7,1.2){${\scriptstyle{10}}$}
\put(-2.0,1.4){${\scriptstyle{5}}$}
\put(-2.5,1.9){${\scriptstyle{2}}$}
\put(-2.9,2.2){${\scriptstyle{1}}$}
\put(-3.6,2.7){${\scriptstyle{0.5}}$}
\put(-4.8,4.0){${\scriptstyle{0.1}}$}
}
\caption{Signal cross section (a), background cross section (b), significance (c), 
and signal to background ratio (d) for $\sqrt{s} = 350\GeV$, and an
integrated luminosity $\mathcal{L}=500\fb^{-1}$ for scenario B: $M_0 =
135\GeV$, $M_{1/2} = 325\GeV$, $A_0 = -135\GeV$, and $\tan\beta = 10$,
see Tables~\ref{tab:scenarios} and~\ref{tab:masses}.  }
\label{fig:scenarioB}
\end{figure}

\begin{figure}[ht]
\setlength{\unitlength}{1cm}
\subfigure[Signal cross section $\sigma(e^+e^- \to \tilde\chi^0_1\tilde\chi^0_1\gamma)$ in fb.\label{fig:sigma500}]
        {\scalebox{0.53}{\includegraphics{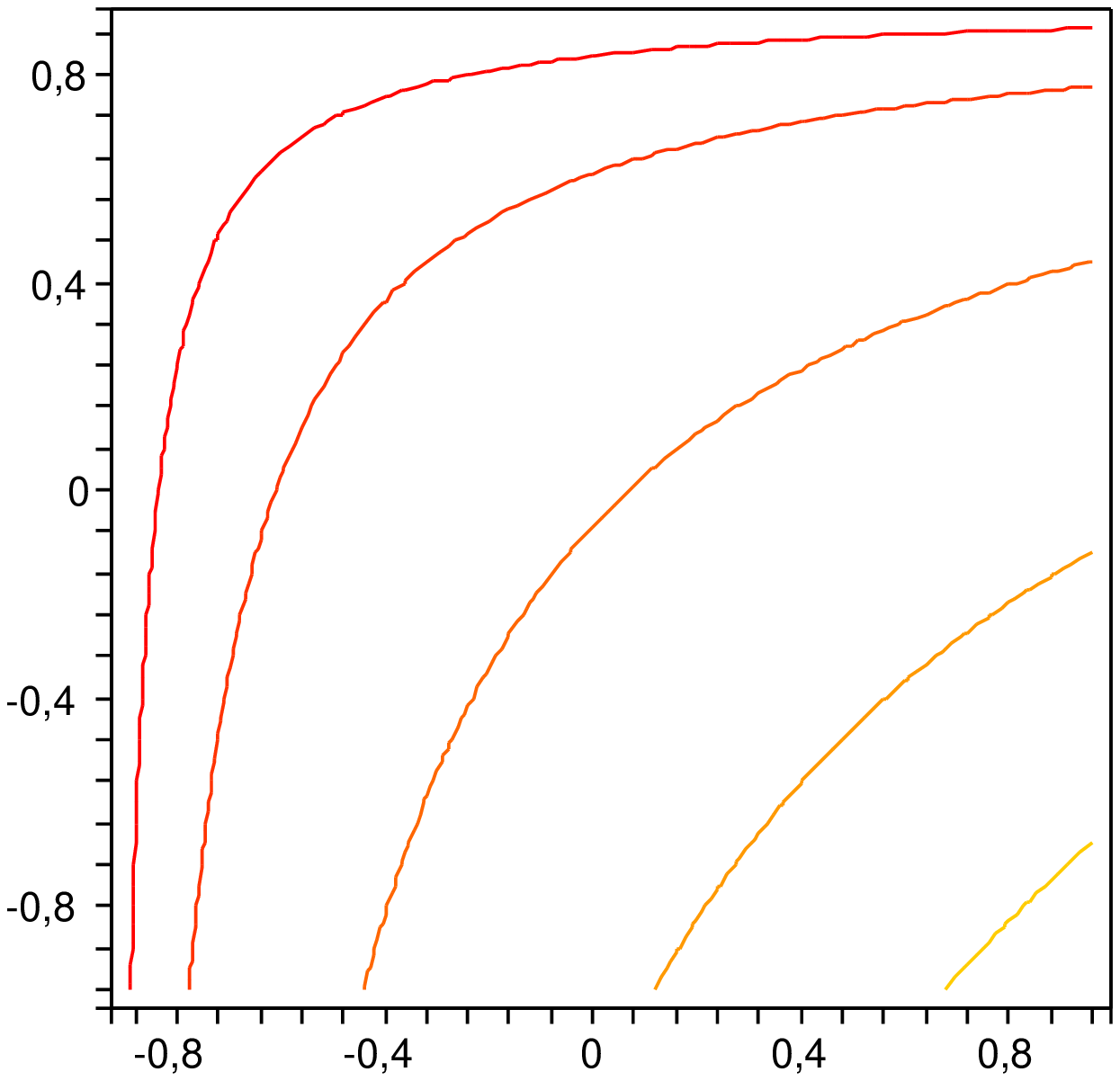}}%
\put(-1.2,0.0){$P_{e^-}$}
\put(-8.0,6.5){$P_{e^+}$}
\put(-1.6,1.3){${\scriptstyle{15}}$}
\put(-2.7,2.4){${\scriptstyle{10}}$}
\put(-4.0,3.7){${\scriptstyle{5}}$}
\put(-5.3,5.1){${\scriptstyle{2}}$}
\put(-6.0,5.8){${\scriptstyle{1}}$}     
}
\hspace{1mm}
\subfigure[Background cross section $\sigma(e^+e^- \to \nu \bar\nu \gamma)$ in fb.\label{fig:back500}]
        {\scalebox{0.53}{\includegraphics{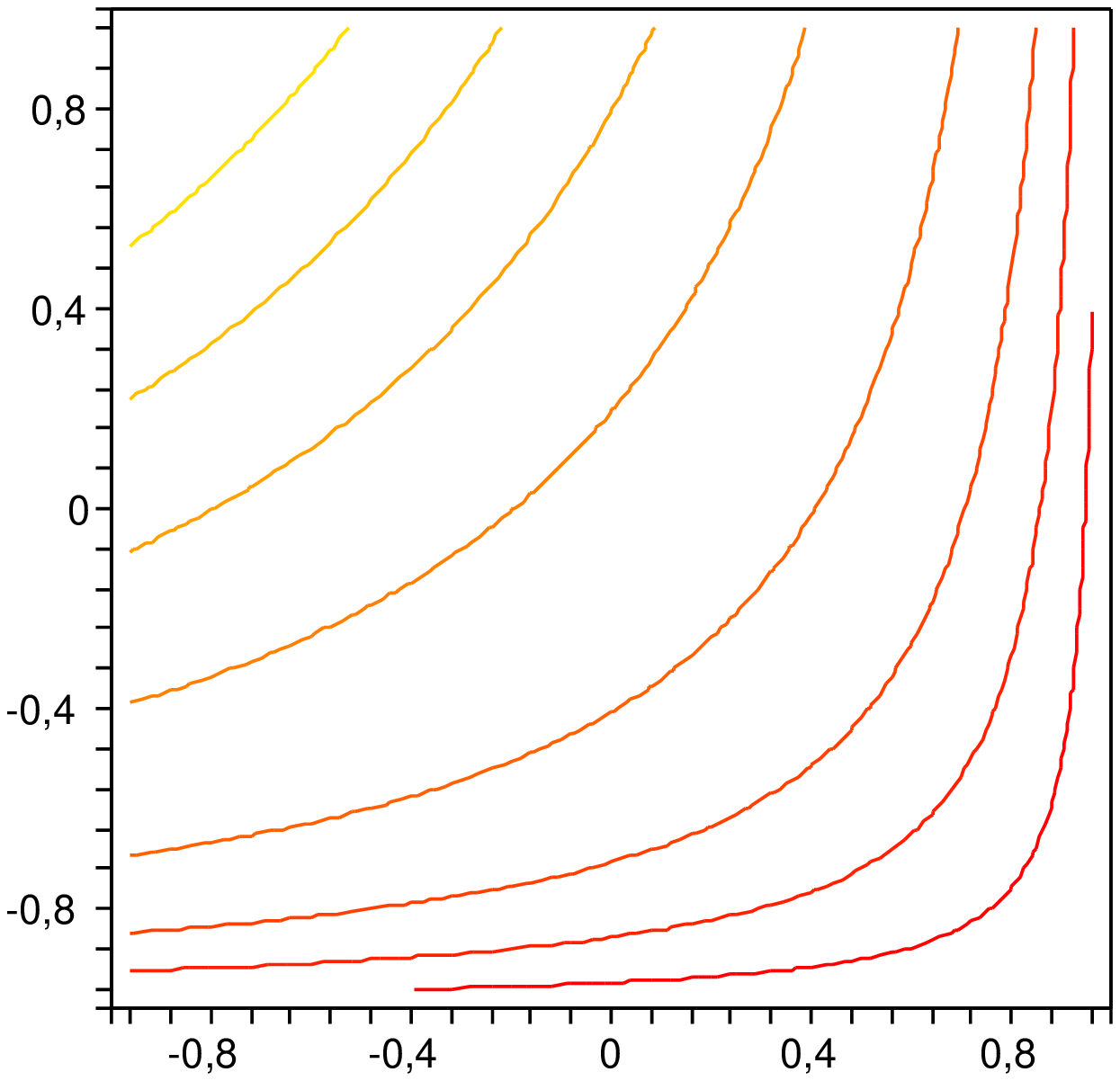}}
\put(-1.2,0){$P_{e^-}$}
\put(-8,6.5){$P_{e^+}$}
\put(-1.8,1.5 ){${\scriptstyle{200}}$}
\put(-2.2,2.0){${\scriptstyle{500}}$}
\put(-2.9,2.5){${\scriptstyle{1000}}$}
\put(-3.6,3.2){${\scriptstyle{2000}}$}
\put(-4.4,4.2){${\scriptstyle{4000}}$}
\put(-5.2,4.9){${\scriptstyle{6000}}$}
\put(-5.8,5.5){${\scriptstyle{8000}}$}
\put(-6.5,6.1){${\scriptstyle{10000}}$}
}

\subfigure[Significance $S$.\label{fig:signi500}]{\scalebox{0.53}{\includegraphics{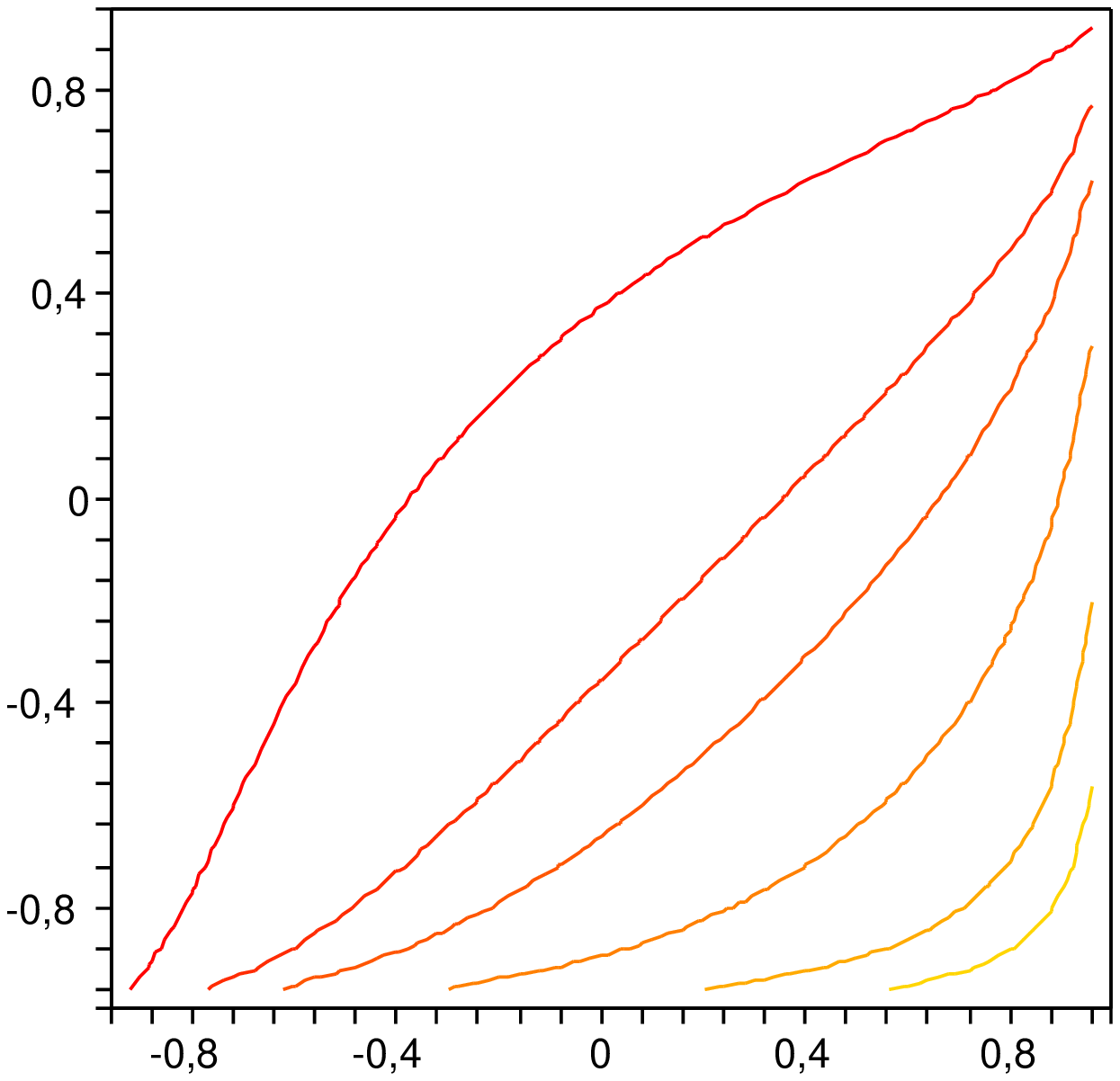}}
\put(-1.2,0){$P_{e^-}$}
\put(-8,6.5){$P_{e^+}$}
\put(-1.3,1.0){${\scriptstyle{30}}$}
\put(-1.7,1.6){${\scriptstyle{20}}$}
\put(-2.4,2.0){${\scriptstyle{10}}$}
\put(-3.1,2.5){${\scriptstyle{5}}$}
\put(-3.8,3.0){${\scriptstyle{3}}$}
\put(-5.0,4.0){${\scriptstyle{1}}$}
}
\hspace{1mm}
\subfigure[Signal to background ratio $r$ in \%.\label{fig:reli500}]{\scalebox{0.53}{\includegraphics{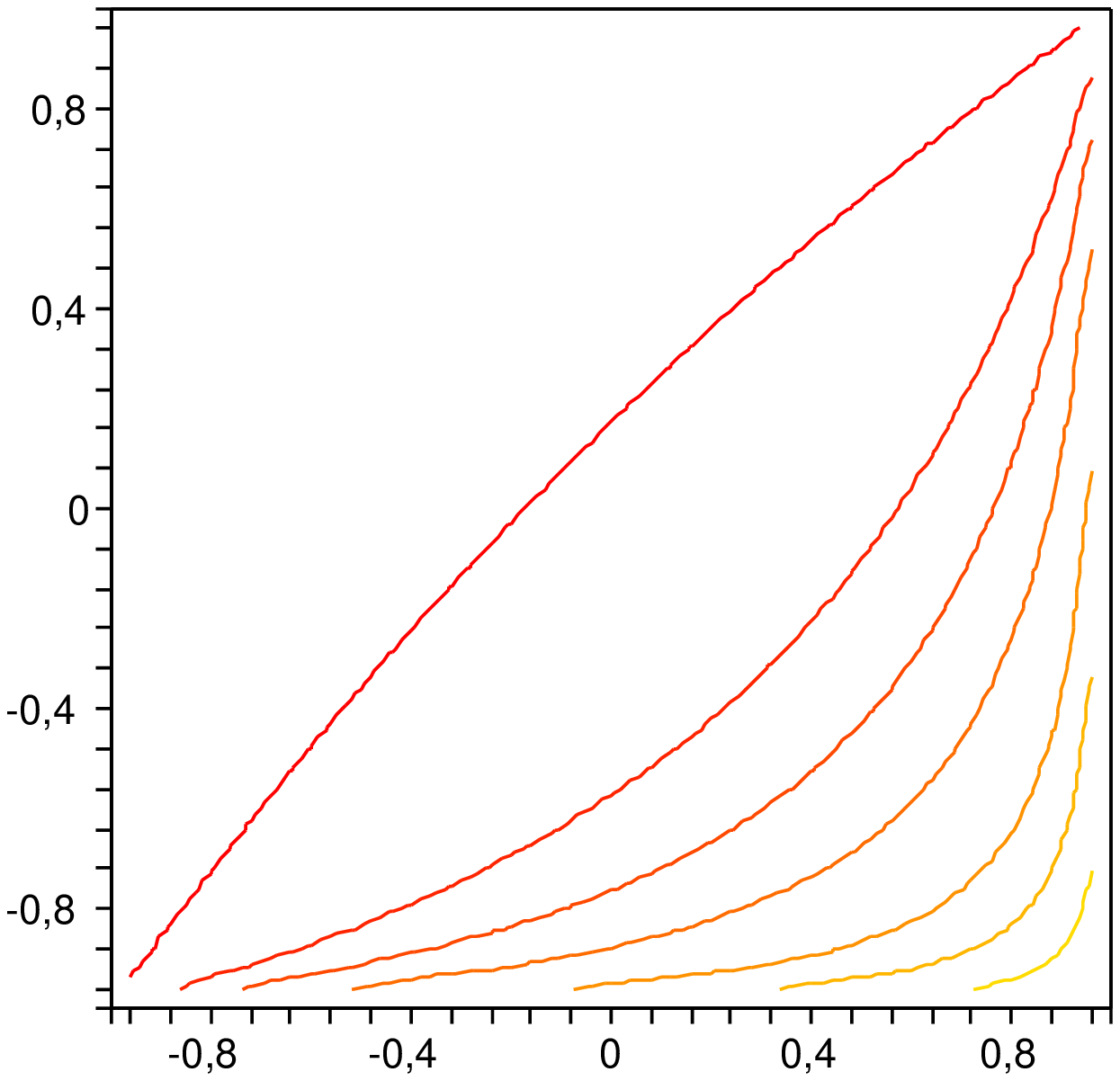}}
\put(-1.2,0){$P_{e^-}$}
\put(-8,6.5){$P_{e^+}$}
\put(-1.2,0.9){${\scriptstyle{20}}$}
\put(-1.7,1.2){${\scriptstyle{10}}$}
\put(-2.0,1.4){${\scriptstyle{5}}$}
\put(-2.4,1.8){${\scriptstyle{2}}$}
\put(-2.8,2.1){${\scriptstyle{1}}$}
\put(-3.4,2.6){${\scriptstyle{0.5}}$}
\put(-4.6,3.8){${\scriptstyle{0.1}}$}
}
\caption{Signal cross section (a), background cross section (b), significance (c),
and signal to background ratio (d) for $\sqrt{s} = 500\GeV$, and an
integrated luminosity$\mathcal{L}=500\fb^{-1}$ for scenario C: $M_0 =
200\GeV$, $M_{1/2} = 415\GeV$, $A_0 = -200\GeV$, and $\tan\beta = 10$,
see Tables~\ref{tab:scenarios} and~\ref{tab:masses}. }
\label{fig:scenarioC}
\end{figure}

Note that the lightest neutralino, $\tilde\chi_1^0$, is mostly bino in
all three scenarios; $98\%$ in scenario A, $99.1\%$ in scenario B, and
$99.5\%$ in scenario C.  Thus in our scenarios, radiative neutralino
production proceeds mainly via right selectron exchange in the $t$ and
$u$ channel. Left selectron exchange and $Z$ boson exchange are
severely suppressed~\cite{Dreiner:2006sb}.  The background process
$e^+e^- \to \nu \bar\nu \gamma$ mainly proceeds via $W$ boson
exchange.  Thus positive electron beam polarization $P_{e^-}>0$ and
negative positron beam polarization $P_{e^+}<0$ should enhance the
signal rate and reduce the background at the same
time~\cite{Choi:1999bs,Dreiner:2006sb}. This effect is clearly
observed in Figs.~\ref{fig:scenarioA}, \ref{fig:scenarioB}, and
\ref{fig:scenarioC} for all scenarios. The signal cross section and
the background vary by more than one order of magnitude over the full
polarization range. 

\medskip

\begin{table}[htb]
\setlength{\belowcaptionskip}{11pt}
\centering
\caption{\small Definition of the mSUGRA scenarios A, B, and C. All values 
are given in GeV. We have fixed $\tan\beta=10$. For
completeness we have included the corresponding value of $\sqrt{s}$
for each scenario.}
\label{tab:scenarios}
\begin{tabular}{cc|ccc|ccc}
\hline 
scenario&$\sqrt{s}$ & $M_0$ & $M_{1/2}$ & $A_0$ & $M_1$ & $M_2$ &
$\mu$ \\[2mm]
\hline 
A & $220$& $90$  & $225$ & $-90$  & $97.5$ &$188$ & $316$ \\[2mm]
B & $350$& $135$ & $325$ & $-135$ & $143$  &$272$ & $444$ \\[2mm]
C & $500$& $200$ & $415$ & $-200$ & $184$  &$349$ & $560$ \\[2mm]
\hline 
\end{tabular}
\end{table}

\begin{table}[htb]
\setlength{\belowcaptionskip}{11pt}
\centering
\caption{\small Spectrum of the lighter SUSY particles for scenarios 
A, B, and C, calculated with SPheno~\cite{Porod:2003um}. All values
are given in GeV. For completeness we have included the corresponding
value of $\sqrt{s}$ for each scenario.}
\label{tab:masses}
\begin{tabular}{cc|ccccccc}
\hline
scenario&$\sqrt{s}$&$m_{\chi_1^0}$ &$m_{\chi_2^0}$ &$m_{\chi_1^\pm}$ & 
 $m_{\tilde{\tau}_1}$   &$m_{\tilde{e}_R}$ & $m_{\tilde{e}_L}$ &$m_{\tilde{\nu}}$ \\[2mm]
\hline 
A & $220$& $92.4$ & $172$ & $172$ & $124$ & $133$ & $189$ & $171$ \\[2mm]
B & $350$& $138$  & $263$ & $263$ & $183$ & $191$ & $270$ & $258$ \\[2mm]
C & $500$& $180$  & $344$ & $344$ & $253$ & $261$ & $356$ & $347$ \\[2mm]
\hline 
\end{tabular}
\end{table} 
\medskip

\begin{table}[t]
\setlength{\belowcaptionskip}{11pt}
\centering
\caption{ Effective polarization $P_{\rm eff}$, 
cross sections $\sigma$, significance $S$, and signal to 
background ratio $r$ for different beam polarizations $(P_{e^+}|
P_{e^-})$ for Scenario A at $\sqrt{s} = 220\GeV$, with
$\mathcal{L}=500\fb^{-1}$.}
\label{tab:xa}
\begin{tabular}{c|cccccc}
\hline
 Scenario A &$(0|0)$&$(0|0.8)$&$(-0.3|0.8)$&$(-0.6|0.8)$&$(0|0.9)$&$(-0.3|0.9)$\\[1mm]
\hline
$P_{\rm eff}=\frac{P_{e^-}-P_{e^+}^{\phantom{k}}}{1-P_{e^-}P_{e^+}}$&$0$&$0.8$&$0.89$&$0.95$&$0.9$&$0.94$\\[2mm]
$\sigma(\signal)$&$6.7\fb$&$12\fb$&$16\fb$&$19\fb$&$13\fb$&$16\fb$\\[1mm]
$\sigma(\Backgroundnu)$& $2685\fb$ & $652\fb$ & $534\fb$ & $416\fb$&$398\fb$&$360\fb$\\[1mm] 
$S$ & $2.9$ &$10$ & $15$ & $20$&$14$& $19$\\[1mm]
$r$ & $0.3\%$ & $1.8\%$ & $2.9\%$ & $4.6\%$&$3.2\%$&$4.6\%$\\[1mm]
\hline 
\end{tabular}
\end{table} 

\begin{table}[ht]
\setlength{\belowcaptionskip}{11pt}
\centering
\caption{ Effective polarization $P_{\rm eff}$,
Cross sections $\sigma$, significance $S$, and signal to 
background ratio $r$ for different beam polarizations $(P_{e^+}|P_{
e^-})$ for Scenario B at $\sqrt{s} = 350\GeV$, with $\mathcal{L}=
500\fb^{-1}$.}
\label{tab:xb}
\begin{tabular}{c|cccccc}
\hline
Scenario B &$(0|0)$&$(0|0.8)$&$(-0.3|0.8)$&$(-0.6|0.8)$&$(0|0.9)$&$(-0.3|0.9)$\\[1mm]
\hline
$P_{\rm eff}=\frac{P_{e^-}-P_{e^+}^{\phantom{k}}}{1-P_{e^-}P_{e^+}}$&$0$&$0.8$&$0.89$&$0.95$&$0.9$&$0.94$\\[2mm]
$\sigma(\signal)$&$5.5\fb$&$9.6\fb$&$13\fb$&$15\fb$&$10.2\fb$&$13.3\fb$\\[1mm]
$\sigma(\Backgroundnu)$& $3064\fb$ & $651\fb$ & $481\fb$ & $312\fb$&$350\fb$&$272\fb$\\[1mm] 
$S$ & $2.2$ &$8.4$ & $13$ & $19$ & $12$ & $18$\\[1mm]
$r$ & $0.2\%$ & $1.5\%$ & $2.6\%$ & $4.9\%$& $2.9\%$ & $4.9\%$\\[1mm]
\hline 
\end{tabular}

\end{table} 
\begin{table}[h!]
\setlength{\belowcaptionskip}{11pt}
\centering
\caption{ Effective polarization $P_{\rm eff}$,
Cross sections $\sigma$, significance $S$, and signal to background ratio $r$ 
for different beam polarizations $(P_{e^+}|P_{e^-})$ for Scenario C at $\sqrt{s} = 500\GeV$, with
$\mathcal{L}=500\fb^{-1}$.}
\label{tab:xc}
\begin{tabular}{c|cccccc}
\hline
Scenario C &$(0|0)$&$(0|0.8)$&$(-0.3|0.8)$&$(-0.6|0.8)$&$(0|0.9)$&$(-0.3|0.9)$\\[1mm]
\hline
$P_{\rm eff}=\frac{P_{e^-}-P_{e^+}^{\phantom{k}}}{1-P_{e^-}P_{e^+}}$&$0$&$0.8$&$0.89$&$0.95$&$0.9$&$0.94$\\[2mm]
$\sigma(\signal)$&$4.7\fb$&$8.2\fb$&$11\fb$&$13\fb$& $8.6\fb$&$11.2\fb$\\[1mm]
$\sigma(\Backgroundnu)$& $3354\fb$ & $689\fb$ & $495\fb$ & $301\fb$& $356\fb$&$263\fb$\\[1mm] 
$S$ & $1.8$ &$7$ & $11$ & $17$& $10$ & $15$\\[1mm]
$r$ & $0.1\%$ & $1.2\%$ & $2.2\%$ & $4.4\%$ & $2.4\%$ & $4.3\%$\\[1mm]
\hline 
\end{tabular}
\end{table} 

For scenario A, we show the beam polarization dependence of the signal
cross section $\sigma(e^+e^- \to \tilde\chi^0_1\tilde\chi^0_1\gamma)$
in Fig.~\ref{fig:sigma220}, and the dependence of the background cross
section $\sigma(e^+e^- \to \nu \bar\nu \gamma)$ in
Fig.~\ref{fig:back220}. In both cases we have implemented the cuts of
Eqs.~(\ref{cuts1}) and (\ref{cuts2}). The contour lines in the $P_{e^-}
$-$P_{e^+}$ plane of the significance $S$, Eq.~(\ref{significance}),
and the signal to background ratio $r$, Eq.~(\ref{sigbck}), are shown
in Figs.~\ref{fig:signi220} and \ref{fig:reli220} respectively. The
results for scenario B are shown in Fig.~\ref{fig:scenarioB}, and
those for scenario C are shown in Fig.~\ref{fig:scenarioC}.

\medskip

In order to quantify the behavior, we give the values for the signal
and background cross sections, the significance $S$ and the signal to
background ratio $r$ for a specific set of beam polarizations $(P_{e^+
}|P_{e^-}) = (0|0)$, $(0|0.8)$, $(-0.3|0.8)$, $(-0.6|0.8)$, $(0|0.9)$,
and $(-0.3|0.9)$ in Tables~\ref{tab:xa}, \ref{tab:xb}, \ref{tab:xc} for
the scenarios A, B, and C, respectively.
In the Tables, we also give the values for the effective beam
polarization 
\begin{equation}
    P_{\rm eff}=\frac{P_{e^-}-P_{e^+}}{1-P_{e^-}P_{e^+}},
\end{equation}
which is a helpful quantity for discussing
combined polarizations when the electron and the positron beam
polarizations have the opposite sign.
For example, a double beam polarization of $P_{e^-} = 0.8$ 
and $P_{e^+} = -0.3$ 
acts effectively as (approximately) $P_{e^-} = 0.89$ of single beam 
polarization~\cite{Omori:1999uz}.
We find that an additional
positron polarization $P_{e^+} = -30\%$ enhances the significance $S$
by factors $\{ 1.5,1.5,1.6\}$ in scenarios $\{{\rm A,B,C}\}$,
respectively, compared to beams with only $e^-$ polarization $(P_{e^+}
|P_{e^-}) =(0|0.8)$, and by factors $\{1.4, 1.5, 1.5\}$ in scenarios
$\{{\rm A,B,C}\}$, respectively, for $(P_{e^+}|P_{e^-}) =(-0.3|0.9)$
compared to $(P_{e^+}|P_{e^-}) =(0|0.9)$.  The signal to background
ratio $r$ is enhanced by $\{ 1.7,1.7,1.8\}$ for $(P_{e^+}|P_{e^-})=
(-0.3|0.8)$ compared to $(P_{e^+}|P_{e^-})= (0|0.8)$ and by
$\{1.4,1.7,1.8\}$ for $(P_{e^+}|P_{e^-})= (-0.3|0.9)$ compared to
$(P_{e^+}|P_{e^-})= (0|0.9)$.  If the positron beams would be
polarized by $P_{e^+} = -60\%$, the enhancement factors for $S$ are
$\{ 2,2.3,2.4\}$, and for $r$ they are $\{ 2.5,3.2,3.6\}$. For
$P_{e^-}= 0.8$, it is only with positron polarization that we obtain
values of $r$ clearly above 1\%. If we have $P_{e^-}= 0.9$, then $r$
exceeds $1\%$ without positron beam polarization.

\medskip

Since the neutralinos are mainly bino, the signal cross section also
depends sensitively on the mass $m_{\tilde{e}_R}$ of the right
selectron.  In scenarios $\{{\rm A,B,C}\}$ the masses are
$m_{\tilde{e}_R}=\{133,191,261\}$~GeV, respectively, see
Table~\ref{tab:masses}.  For larger masses, the signal to background
ratio drops below $r<1\%$. With $(P_{e^+}|P_{e^-}) =(-0.3|0.8)$, this
happens for $m_{\tilde{e}_R}=\{214,300,390\}$~GeV, and the
significance would be $S<5$. These selectron masses correspond to the
mSUGRA parameter $M_0=\{190,270,350\}$~GeV.

\subsection*{Note on Wino- and Higgsino-like neutralinos\label{higgsinowino}}

In anomaly mediated supersymmetry breaking (AMSB) scenarios,
the gaugino mass parameters are roughly related by
$M_1 \approx 2.8 M_2$ at the weak scale~\cite{Drees:2004jm}.
As a consequence, the lightest neutralino is wino like,
and left slepton exchange is enhanced in the reaction
$e^+e^- \to \tilde\chi^0_1\tilde\chi^0_1\gamma$.
Therefore, it is no longer possible to increase the signal and
to reduce the background simultaneously by a suitable choice of
beam polarizations.
In AMSB scenarios, also $m_0\gsim 400$~GeV is bounded from 
below~\cite{Datta:2001er} and rather large. For smaller $m_0$,
either the lightest stau is the
LSP or sleptons would already have been observed.   
The relatively heavy selectrons lead then
to a suppression
of the cross section for radiative neutralino production.
For example in the scenario
SPS9~\cite{Aguilar-Saavedra:2005pw,Allanach:2002nj}
with  $m_0= 450$~GeV and 
$m_{\tilde e_L} = 387$~GeV,  $m_{\tilde e_R}=375$~GeV,
$m_{\tilde \tau_1} = 349$~GeV,
calculated with  SPheno~\cite{Porod:2003um},
we find that the significance is $S<1$ and the signal to background
ratio $r<0.03\%$, for various values of the beam polarizations
at $\sqrt s=500$~GeV.
Finally, in AMSB scenarios the lightest neutralino
$\tilde\chi^0_1$ is nearly mass degenerate with
the lightest chargino $\tilde\chi^\pm_1$. 
Thus, if  the radiative production of neutralinos 
is kinematically accessible,
also the pair production of charginos
$e^+e^- \to \tilde\chi^+_1\tilde\chi^-_1$
can be well observed.
In this context it is interesting to note 
that the radiative production of charginos
 $e^+e^- \to \tilde\chi^+_1\tilde\chi^-_1\gamma$
might be important, 
if the charginos decay
almost invisibly~\cite{Datta:2001zw}.
This could happen,
if the neutralino-chargino mass 
difference is such,
that the charginos decay
into rather soft pions
$\tilde\chi^\pm_1\to\tilde\chi^0_1\pi^\pm$.

\medskip

The cross section of radiative neutralino production 
is also small in models, where the lightest neutralino is
higgsino-like, \textit{i.e.} $\mu \ll M_2 $. 
For example, for $\mu<300$~GeV and $M_2>500$~GeV
we find $S<1$ and 
$\sigma(e^+e^- \to \tilde\chi^0_1\tilde\chi^0_1\gamma)<1$~fb,
since the gaugino couplings of the neutralino are suppressed.
The couplings of the neutralino to the $Z$-boson 
are proportional to the squared difference
of the higgsino couplings, $|N_{13}|^2-|N_{14}|^2$
see Ref.~\cite{Dreiner:2006sb},
and thus too small to lead to an observable signal.

\section{Summary and Conclusions\label{Summary}}

We have studied radiative neutralino production $e^+e^- \to\tilde\chi
^0_1\tilde\chi^0_1\gamma$ at the ILC with longitudinally polarized
beams. For the center-of-mass energies $\sqrt{s} = 220\GeV$, $350\GeV
$, and $500\GeV$, we have considered three specific mSUGRA inspired
scenarios. In our scenarios, only radiative neutralino production is
kinematically accessible, since the other supersymmetric particles are
too heavy to be pair produced. We have investigated the beam
polarization dependence of the cross section from radiative neutralino
production and the background form radiative neutrino production
$e^+e^- \to \nu\bar\nu \gamma$.

\medskip

We have shown that polarized beams enhance the signal and suppress the
background simultaneously and significantly. In our scenarios, the
signal cross section for $(P_{e^+}|P_{e^-}) = (-0.3|0.8)$ is larger
than $10\fb$, the significance $S>10$, and the signal to background
ratio is about $2-3\%$. The background cross section can be reduced to
$500\fb$.  Increasing the positron beam polarization to $P_{e^+} =
-0.6$, both the signal cross section and the significance increase by
about $25\%$, in our scenarios. For $(P_{e^+}|P_{e^-}) = (0.0|0.9)$
the radiative neutralino production signature is observable at the ILC
but both the significance and the signal to background ratio are
considerable improved for $(P_{e^+}|P_{e^-}) = (-0.3|0.9)$, making
more detailed investigations possible. The electron and positron beam
polarization at the ILC are thus essential tools to observe radiative
neutralino production. For unpolarized beams this process cannot be
measured.

\medskip

We conclude that radiative neutralino production can and should 
be studied at $\sqrt{s} =500\GeV$, as well as at the lower energies
$\sqrt{s} = 220\GeV$ and  $\sqrt{s} =350\GeV$,
which are relevant for Higgs and top physics.
We have shown that for these energies there are scenarios, 
where other SUSY particles 
like heavier neutralinos, charginos and sleptons are too heavy to
be pair produced. 
In any case, a pair of radiatively produced neutralinos
is the lightest accessible state of SUSY particles to be produced at 
the linear collider.

\section*{Acknowledgments}
\noindent

We would like to thank Gudrid Moortgat-Pick and Klaus Desch for very
helpful discussions. This work was supported by the SFB Transregio
33: The Dark Universe. One of us (OK) would like to thank the organizers 
of the ECFA-ILC workshop
Valencia 06, where the initial idea for this work was developed.

\end{document}